\documentclass[twocolumn,journal]{IEEEtran}
\usepackage[T1]{fontenc}
\usepackage[latin9]{inputenc}
\usepackage{amsmath}
\usepackage{amssymb}
\usepackage{graphicx}
\usepackage[unicode=true,
 bookmarks=true,bookmarksnumbered=true,bookmarksopen=true,bookmarksopenlevel=1,
 breaklinks=false,pdfborder={0 0 0},pdfborderstyle={},backref=false,colorlinks=false]
 {hyperref}
\hypersetup{pdftitle={Enhanced Sensitivity of Degenerate System Made of Two Unstable Resonators Coupled by Gyrator Operating at an Exceptional Point},
 pdfauthor={Kasra Rouhi},
 pdfpagelayout=OneColumn, pdfnewwindow=true, pdfstartview=XYZ, plainpages=false}

\usepackage{fancyhdr}

\makeatletter
\let\oldforeign@language\foreign@language
\DeclareRobustCommand{\foreign@language}[1]{%
  \lowercase{\oldforeign@language{#1}}}

\usepackage[caption=false,font=footnotesize]{subfig}

\makeatother

\begin{document}
\title{Enhanced Sensitivity of Degenerate System Made of Two Unstable Resonators Coupled by Gyrator Operating at an Exceptional Point}
\author{Kasra Rouhi, Alireza Nikzamir, Alexander Figotin, and Filippo Capolino\thanks{Kasra Rouhi, Alireza Nikzamir, and Filippo Capolino are with the Department
of Electrical Engineering and Computer Science, University of California,
Irvine, CA 92697 USA, e-mails: \protect\href{mailto:kasra.rouhi@uci.edu}{kasra.rouhi@uci.edu},
\protect\href{mailto:anikzami@uci.edu}{anikzami@uci.edu}, and \protect\href{mailto:f.capolino@uci.edu}{f.capolino@uci.edu}.}\thanks{Alexander Figotin is with the Department of Mathematics, University
of California, Irvine, CA 92697 USA, e-mail: \protect\href{mailto:afigotin@uci.edu}{afigotin@uci.edu}.}}
\maketitle

\thispagestyle{fancy}

\begin{abstract}
We demonstrate that a circuit comprising two unstable LC resonators
coupled via a gyrator supports an exceptional point of degeneracy
(EPD) with purely real eigenfrequency. Each of the two resonators
includes either a capacitor or an inductor with a negative value,
showing purely imaginary resonance frequency when not coupled to the
other via the gyrator. With external perturbation imposed on the system,
we show analytically that the resonance frequency response of the
circuit follows the square-root dependence on perturbation, leading
to possible sensor applications. Furthermore, the effect of small
losses in the resonators is investigated, and we show that losses
lead to instability. In addition, the EPD occurrence and sensitivity
are demonstrated by showing that the relevant Puiseux fractional power
series expansion describes the eigenfrequency bifurcation near the
EPD. The EPD has the great potential to enhance the sensitivity of
a sensing system by orders of magnitude. Making use of the EPD in
the gyrator-based circuit, our results pave the way to realize a new
generation of high-sensitive sensors to measure small physical or
chemical perturbations.
\end{abstract}


\IEEEpeerreviewmaketitle{}

\section{Introduction}

\IEEEPARstart{A}{n} exceptional point of degeneracy (EPD) is a point
in parameter space at which the eigenmodes of the circuit, namely
the eigenvalues and the eigenvectors, coalesce simultaneously \cite{heiss2000repulsion,Figotin2003Oblique,figotin2005gigantic,ramezani2012exceptional}.
As the remarkable feature of an EPD is the strong full degeneracy
of at least two eigenmodes, as mentioned in \cite{berry2004Physics},
the significance of referring to it as a \textquotedblleft degeneracy\textquotedblright{}
is here emphasized, hence including \textquotedbl D\textquotedbl{}
in EPD. An EPD in the system is reached when the system matrix is
similar to a matrix that contains a non-trivial Jordan block. EPD-induced
sensitivity according to the concept of parity-time (PT) symmetry
in multiple coupled resonators has been studied \cite{Bender1998Real,Hodaei2014Parity,Chen2018Generalized}.
Also, the electronic circuits with EPD based on PT symmetry have been
expressed in \cite{stehmann2004observation,Schindler2011Experimental}
and then more developed in \cite{sakhdari2018ultrasensitive,zhang2019noninvasive}
where the circuits are made of two coupled resonators with gain-loss
symmetry and a proper combination of parameters leads to an EPD. Primarily,
it has been confirmed that the eigenvalues bifurcation feature at
EPD can significantly increase the effect of external perturbation;
namely, the sensitivity of resonance frequency to components value
perturbations can be enhanced. Moreover, frequency splitting happens
at degenerate frequencies of the system where eigenmodes coalesce,
and this feature at EPDs has been investigated to conceive a new generation
of sensors \cite{wiersig2014enhancing,Wiersig2016Sensors,wiersig2020robustness,wiersig2020review}.
The resulting perturbation leads to a shift in the system resonance
frequency that can be recognized and measured using the proper measurement
setup \cite{wiersig2014enhancing}. When a second-order EPD at which
specifically two eigenstates coalesce is subjected to a small external
perturbation, the resulting eigenvalue splitting is proportional to
the square root of external perturbation value, which is bigger than
the case of linear splitting for conventional degeneracies \cite{wiersig2020prospects}.
The concept of EPD has been employed in various sensing schemes such
as optical microcavities \cite{Chen2018Generalized}, optical microdisk
\cite{gwak2021rayleigh}, electron beam devices \cite{rouhi2021exceptional},
mass sensors \cite{djorwe2019exceptional}, and bending curvature
sensors \cite{wang2018review}.

The gyrator is a two-terminal element with the property of transmission
phase shift in one direction differs by $\pi$ from that for transmission
in the other direction \cite{tellegen1948gyrator}. Another property
of the gyrator network is that of impedance inversion. The inductance
at the output of the gyrator is observed as capacitance at the input
port, and a voltage source is transformed to a current source. A relevant
alias for the gyrator might be the \textquotedblleft dualizer\textquotedblright{}
since it can interchange current and voltage roles and turns an impedance
into its dual \cite{hamill1993lumped}. Gyrators could be designed
directly as integrated circuits \cite{sheahan1966integratable,rao1966direct}.
Also, many operational-amplifier (opamp) gyrator circuits have been
proposed \cite{morse1964gyrator,riordan1967simulated,antoniou1967gyrators},
which can be classified into two types. First, 3-terminal gyrator
circuits in which both ports are grounded \cite{morse1964gyrator};
second, 4-terminal gyrator circuits in which the output port is floating
\cite{riordan1967simulated,antoniou1967gyrators}. Because of the
availability of different realizable circuits for gyrators and their
versatility for practical circuit devices, gyrator-based circuits
may form an essential part of integrated circuit technology in a wide
range of applications.

In this paper, we study the second-order EPDs in a gyrator-based sensing
circuit as Fig. \ref{Fig: Circuit} and explore its enhanced sensitivity
(variation in sensor's resonance frequencies to external perturbations)
and its potentials for sensing devices in the vicinity of EPD. Two
series LC resonators are coupled in the utilized circuit via an ideal
gyrator, as explained in \cite{nikzamir2021demonstration}. Contrary
to the study in \cite{nikzamir2021demonstration}, this paper demonstrates
the conditions to get the EPD with real eigenfrequency by using unstable
resonators. In other words, we study the case of two unstable resonators
coupled via an ideal gyrator. A general mathematical approach for
constructing lossless circuits for any conceivable Jordan structure
has been developed in \cite{figotin2020synthesis}, including the
simplest possible circuit as in Fig. \ref{Fig: Circuit} and other
circuits related to the Jordan blocks of higher dimensions. In addition,
important issues related to operational stability, perturbation analysis,
and sensitivity analysis are studied in \cite{figotin2021perturbations},
whereas analysis of stability or instability by adding losses to the
circuit is not discussed. We show that the gyrator-based circuit can
achieve EPD with real eigenfrequency even when two unstable resonators
are used in the circuit. Hence, dispersion diagrams correspond to
perturbation in the circuit's parameters show the eigenfrequencies
split. Then, we show examples for different cases and analyze the
output signal by using time-domain simulations. We then study the
impact of small losses in the circuit and explain how they can make
it unstable. Besides, we look at the sensitivity of circuit eigenfrequencies
to component variations, and we show that the Puiseux fractional power
series expansion well approximates the bifurcation of the eigenfrequency
diagram near the EPD \cite{Kato1995Perturbation}. The sensitivity
enhancement is attributed to the second root topology of the eigenvalues
in parameter space, peculiar to the second-order EPD. Lastly, we examine
the gyrator-based circuit's enhanced sensitivity and provide a practical
scenario to detect physical parameters variations and material characteristics
changes. Besides exploring EPD physics in the gyrator-based circuits,
this work is also important for understanding the instability in resonators.
The given analysis and circuit show promising potential in the novel
ultra high-sensitive sensing applications.

\section{Gyrator Characteristic}

A gyrator is a two-port component that couples an input port to an
output port by a gyration resistance value. It is a lossless and storage-less
two-port network that converts circuits at the gyrator output into
their dual, with respect to the gyration resistance value \cite{ehsani1993power}.
For instance, this component can make a capacitive circuit behave
inductively, a series LC resonator behave like a parallel LC resonator,
and so on. This device allows network realizations of two-port devices,
which cannot be realized by just the basic components, i.e., resistors,
inductors, capacitors, and transformers. In addition, the gyrator
could be considered a more fundamental circuit component than the
ideal transformer because an ideal transformer can be made by cascading
two ideal gyrators, but a gyrator cannot be made from transformers
\cite{tellegen1948gyrator}. The circuit symbol for the ideal gyrator
is represented in Fig. \ref{Fig: Circuit} (red dashed box), and the
defining equations are \cite{tellegen1948gyrator,shenoi1965practical}

\begin{equation}
\begin{cases}
v_{2}=R_{g}i_{1}\\
v_{1}=-R_{g}i_{2}
\end{cases}
\end{equation}
where $R_{g}$ is called gyration resistance and has a unit of Ohm.
A gyrator is a nonreciprocal two-port network represented by an asymmetric
impedance matrix $\mathbf{\underline{Z}}$ as \cite{shenoi1965practical}

\begin{equation}
\mathbf{\underline{Z}}=\left[\begin{array}{cc}
0 & -R_{g}\\
R_{g} & 0
\end{array}\right].
\end{equation}

\section{EPD Condition in The Lossless Gyrator-Based Circuit\label{sec:Lossless-EPD-Condition}}

\begin{figure}[t]
\centering{}\includegraphics[width=3.5in]{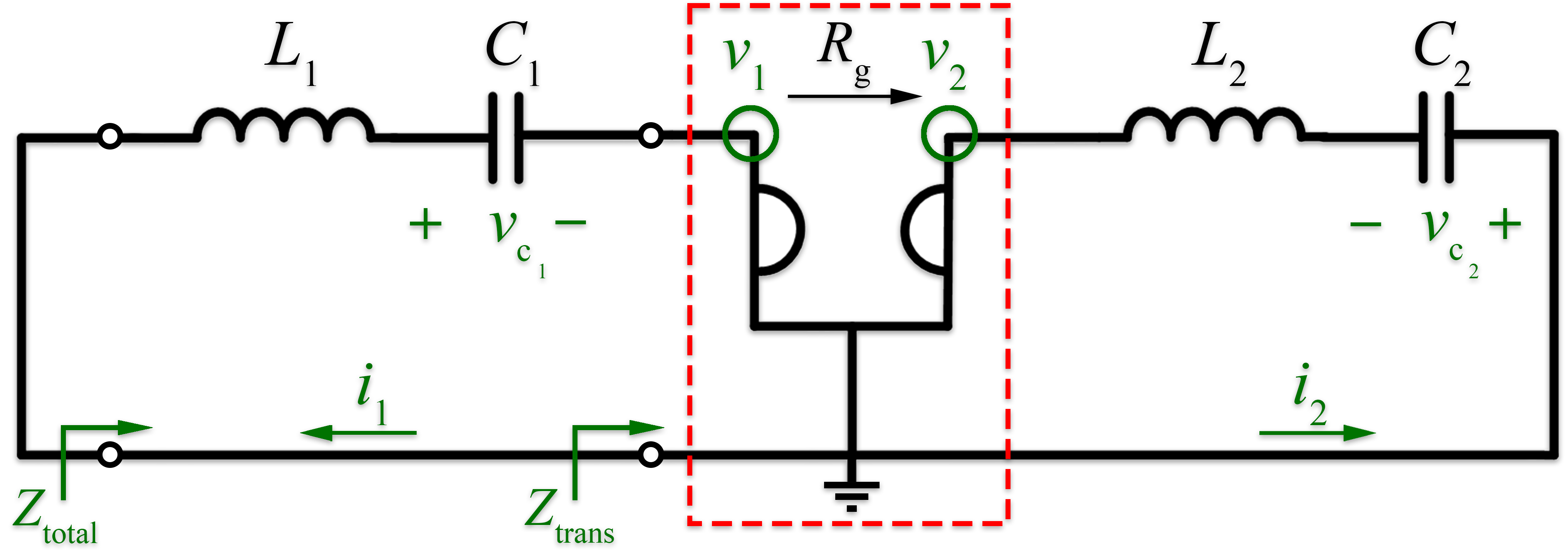}\caption{The schematic illustration of the proposed gyrator-based circuit with
the ideal gyrator indicated by the red dashed box. In this circuit,
two different (unstable) LC resonators are embedded in a series configuration,
coupled via an ideal gyrator.\label{Fig: Circuit}}
\end{figure}

This section provides an analysis of a gyrator-based circuit in which
two series LC resonators are coupled via an ideal gyrator as illustrated
in \ref{Fig: Circuit}. We first assume that all components are ideal,
and the circuit does not contain any resistance. The circuit resembles
the one in \cite{nikzamir2021demonstration}, but here the two resonance
angular frequencies $\omega_{01}=1/\sqrt{C_{1}L_{1}}$, and $\omega_{02}=1/\sqrt{C_{2}L_{2}}$
of the two uncoupled resonators are imaginary with a negative sign
(also the counterpart with the positive sign is a resonance), since
we consider three cases: (i) both $L_{1}$ and $L_{2}$ are negative
while the capacitors have a positive value, (ii) both $C_{1}$ and
$C_{2}$ are negative while the inductors have a positive value, and
(iii) $L_{1}$($C_{1}$) and $C_{2}$($L_{2}$) are negative while
other elements have a positive value. Then we investigate the possibility
of the occurring EPD in the cases just mentioned. In the past years,
EPDs have been found by using balanced loss and gain in a PT symmetry
scheme \cite{Heiss2004Exceptional,Schindler2011Experimental,Chen2018Generalized}.
More recently, EPDs have also been found in systems with time-periodic
modulation \cite{Kazemi2019Exceptional,kazemi2020ultra}. Here, we
obtain EPDs by using the negative inductance and capacitance in the
gyrator-based circuit, constituting a new class of EPD-based circuits.

We consider the Kirchhoff voltage law equations in the time-domain
for two loops of the circuit in Fig. \ref{Fig: Circuit}. In order
to find the solution of the circuit differential equations, it is
convenient to define the state vector as $\boldsymbol{\Psi}(t)\equiv[Q_{1},Q_{2},\dot{Q}_{1},\dot{Q}_{2}]^{T}$,
where $T$ denotes the transpose operator. The state vector consists
of stored charges in the capacitors $Q_{n}=\int i_{n}dt=C_{n}v_{c_{n}}$,
and their time derivative (currents) $\dot{Q}_{n}=i_{n}$, $n=1,\ldots,4.$
We utilize then the Liouvillian formalism for this circuit as \cite{nikzamir2021demonstration}

\begin{equation}
\mathcal{\mathrm{\frac{d\boldsymbol{\Psi}(t)}{dt}=\mathbf{\underline{M}}\boldsymbol{\Psi}(t),\;}\mathbf{\underline{M}}}=\left(\begin{array}{cccc}
0 & 0 & 1 & 0\\
0 & 0 & 0 & 1\\
-\omega_{01}^{2} & 0 & 0 & \frac{R_{g}}{L_{1}}\\
0 & -\omega_{02}^{2} & -\frac{R_{g}}{L_{2}} & 0
\end{array}\right),\label{eq: eigenvalue problem}
\end{equation}
where $\underline{\mathbf{M}}$ is the $4\times4$ circuit matrix.
Assuming time harmonic dependence of the form $Q_{n}\varpropto e^{j\omega t}$,
we obtain the characteristic equation allowing us to find the eigenfrequencies
by solving $\det(\underline{\mathbf{M}}-j\omega\underline{\mathbf{I}})=0$,
where $\underline{\mathbf{I}}$ is the identity matrix. The corresponding
characteristic equation of the circuit is

\begin{equation}
\omega^{4}-\omega^{2}\left(\omega_{01}^{2}+\omega_{02}^{2}+\frac{R_{g}^{2}}{L_{1}L_{2}}\right)+\omega_{01}^{2}\omega_{02}^{2}=0,\label{eq:Characteristic}
\end{equation}
where any solution $\omega$ is an eigenfrequency of the circuit.
In the case of $R_{g}=0$, the two resonators are uncoupled, and the
circuit has two eigenfrequency pairs of $\omega_{1,3}=\pm\omega_{01}$,
and $\omega_{2,4}=\pm\omega_{02},$ that are purely imaginary (in
contrast to the case studies in \cite{nikzamir2021demonstration},
where the resonance frequencies have real values). All the $\omega$'s
coefficients of the characteristic equation are real, so $\omega$
and $\omega^{*}$ are both roots of the characteristic equation, where
{*} indicates the complex conjugate operator. Moreover, it is a quadratic
equation in $\omega^{2}$; therefore, $\omega$ and $-\omega$ are
both solutions of the Eq. (\ref{eq:Characteristic}). As we mentioned
before, we only consider unstable resonators, i.e., resonators with
imaginary resonance frequency. Therefore, only one circuit element
in each resonator should have a negative value, leading to $\omega_{01}^{2}$
and $\omega_{02}^{2}$ with negative values. After finding the solutions
of the characteristic equation, the angular eigenfrequencies (resonance
frequencies) of the circuit are expressed as

\begin{equation}
\omega_{1,3}=\pm\sqrt{a+b},\;\omega_{2,4}=\pm\sqrt{a-b},\label{eq: Eigenfrequencies}
\end{equation}
where

\begin{equation}
a=\frac{1}{2}\left(\omega_{01}^{2}+\omega_{02}^{2}+\omega_{g}^{2}\right),\label{eq: EPD_a}
\end{equation}

\begin{equation}
\begin{array}{c}
b^{2}=a^{2}-\omega_{01}^{2}\omega_{02}^{2}\end{array},\label{eq: EPD_b}
\end{equation}
where it has been convenient to define $\omega_{g}^{2}=R_{g}^{2}/(L_{1}L_{2})$,
that may be positive or negative depending on the considered case.
According to Eq. (\ref{eq: Eigenfrequencies}), the EPD condition
requires

\begin{equation}
b=0,\label{eq: EPDCondition}
\end{equation}
leading to an EPD angular frequency $\omega_{\mathit{e}}=\sqrt{a}$
(with its negative pair $-\omega_{\mathit{e}}$). According to Eq.
(\ref{eq: EPD_b}), the EPD condition is rewritten as $a^{2}=\omega_{01}^{2}\omega_{02}^{2}.$
As in \cite{nikzamir2021demonstration}, we consider positive values
for $a$ to have a real EPD angular frequency $\omega_{\mathit{e}}$
, so we have

\begin{equation}
\omega_{01}^{2}+\omega_{02}^{2}+\omega_{g}^{2}>0.\label{eq:EPD_Cond_real}
\end{equation}

Finally, the EPD frequency is calculated by using Eqs. (\ref{eq: EPD_a}),
(\ref{eq: EPD_b}), and (\ref{eq: EPDCondition}) as

\begin{equation}
\begin{array}{c}
\omega_{\mathit{e}}=\sqrt{\frac{1}{2}\left(\omega_{01}^{2}+\omega_{02}^{2}+\omega_{g}^{2}\right)}.\end{array}\label{eq: EPDFrequency}
\end{equation}

The last equation can also be rewritten as $\omega_{\mathit{e}}=\sqrt[4]{\omega_{01}^{2}\omega_{02}^{2}}$,
with the quartic square root defined by taking the positive value;
in other words, if we consider that the two unstable frequencies have
the following purely imaginary expression, $\omega_{01}=-j/\sqrt{|C_{1}L_{1}|}$
and $\omega_{02}=-j/\sqrt{|C_{2}L_{2}|}$, the EPD frequency can be
expressed as $\omega_{\mathit{e}}=\sqrt{-\omega_{01}\omega_{02}}$.
We obtain the desired value of a real EPD frequency by optimizing
the values of the components in the circuit. Theoretically, the utilized
optimization method is not critical, and merely we need to found the
proper solution for Eq. (\ref{eq: EPDCondition}). Obviously, practical
limitations also affect the selection of suitable constraints for
optimization. In the particular case the two circuits are identical,
one has $\omega_{0}^{2}\equiv\omega_{01}^{2}=\omega_{02}^{2}=1/(LC)<0$,
and the EPD condition (\ref{eq: EPDCondition}) reduces to $4\omega_{0}^{2}=-\omega_{g}^{2},$that
in turns leads to the EPD angular frequency $\begin{array}{c}
\omega_{\mathit{e}}=\sqrt{-\omega_{0}^{2}}\end{array}$. In the following subsections, we analyze the circuit in three different
cases, i.e., the three different assumptions mentioned earlier.

\subsection{Negative Inductances $L_{1}$ and $L_{2}$\label{subsec:Negative-Inductances}}

\begin{figure*}[tbh]
\centering{}\includegraphics[width=7in]{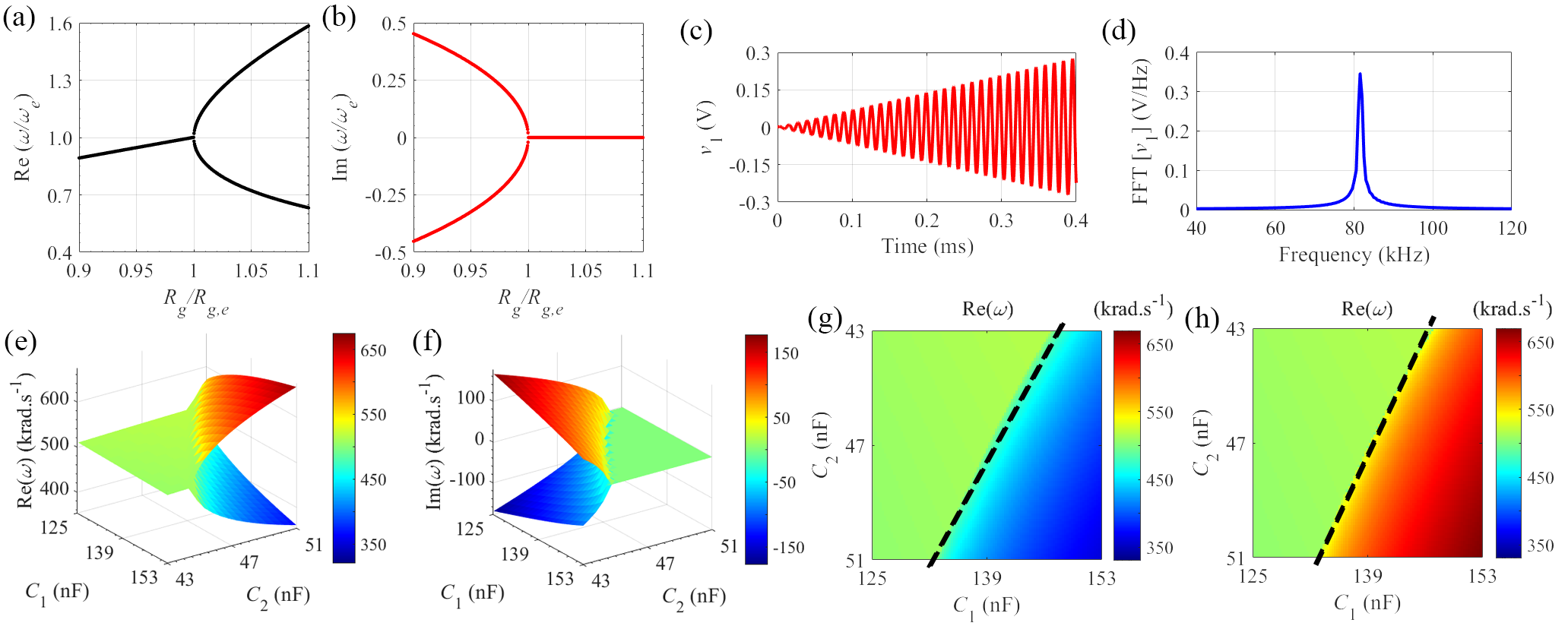}\caption{The sensitivity of the (a) real and (b) imaginary parts of the eigenfrequencies
to gyration resistance perturbation. Voltage $v_{1}(t)$ under the
EPD condition in the (c) time-domain, and (d) frequency-domain. The
frequency-domain result is calculated from $40\mathrm{\:kHz}$ to
$120\:\mathrm{kHz}$ by applying an FFT with $10^{6}$ samples in
the time window of $0\:\mathrm{ms}$ to $0.4\:\mathrm{ms}$. The three-dimensional
plot of the (e) real and (f) imaginary parts of the eigenfrequencies
to $C_{1}$ and $C_{2}$ perturbation. The real part of eigenfrequencies
for (g) higher and (h) lower value of resonance frequencies which
colormap show the resonance frequency value. The black dashed line
in these plots shows the EPD.\label{Fig: EPD-L-Neg}}
\end{figure*}
As a first case, we consider a negative value for both inductances
and a positive value for both capacitances; hence, in this case $\omega_{g}^{2}>0$.
According to the required condition for EPD expressed in Eq. (\ref{eq: EPDCondition})
and by using Eq. (\ref{eq: EPD_b}), the first and second terms in
Eq. (\ref{eq: EPD_a}) are negative and the third term is positive.
Eq. (\ref{eq: EPDFrequency}) shows that, if $\left|\omega_{01}^{2}+\omega_{02}^{2}\right|<\omega_{g}^{2}$
we obtain a real value for EPD frequency, and if $\left|\omega_{01}^{2}+\omega_{02}^{2}\right|>\omega_{g}^{2}$,
the EPD frequency yields an imaginary value.

We explain the procedure to obtain an EPD in this circuit by presenting
an example. Various combinations of values for the circuit's components
can satisfy the EPD condition demonstrated in Eq. (\ref{eq: EPDCondition}),
and here as an example, we consider this set of values: $L_{1}=-47\mathrm{\:\mu H}$,
$L_{2}=-47\mathrm{\:\mu H}$, $C_{2}=47\mathrm{\:nF}$, and $R_{g}=50\:\Omega$.
Then, the capacitance of the first resonator is determined by solving
the resulting quadratic equation from the EPD condition demonstrated
in Eq. (\ref{eq: EPDCondition}). In this example, we consider $C_{1}$
as a sensing capacitance of the circuit, which has a positive value
and it can detect variations in environmental parameters and transform
them into electrical quantities. According to Eq. (\ref{eq: EPDCondition}),
after solving the quadratic equation, two different values for capacitance
in the first resonator are calculated, and we consider $C_{1,\mathit{e}}=139.17\:\mathrm{nF}$
for the presented example. According to the calculated values for
components, both $\omega_{01}^{2}$ and $\omega_{02}^{2}$ have negative
values, with $\omega_{01}=-j391\:\mathrm{krad/s}$, and $\omega_{02}=-j672.82\:\mathrm{krad/s}$,
leading to a positive result for $a$ in Eq. (\ref{eq: EPD_a}) and
real EPD angular frequency $\omega_{\mathit{e}}=512.9\mathrm{\:krad/s}$.
The results in Figs. \ref{Fig: EPD-L-Neg}(a), and (b) show the real
and imaginary parts of perturbed eigenfrequencies obtained from the
eigenvalue problem when $R_{g}$ of the ideal gyrator is perturbed,
revealing the high sensitivity to perturbations.

To investigate the time-domain behavior of the circuit under EPD conditions,
we used the Keysight Advanced Design System (ADS) circuit simulator.
The transient behavior of the coupled resonators with the ideal gyrator
is simulated using the time-domain solver with an initial condition
$v_{c_{1}}(0)=1\:\mathrm{mV}$, where $v_{c_{1}}(t)$ is the voltage
of the capacitor in the left resonator. Fig. \ref{Fig: EPD-L-Neg}(c)
shows the time-domain simulation results of the voltage $v_{1}(t)$,
where $v_{1}(t)$ is the voltage at the gyrator input port (see Fig.
\ref{Fig: Circuit}). The extracted result is obtained in the time
span of $0$ ms to $0.4$ ms. The solution of the eigenvalue problem
in the Eq. (\ref{eq: eigenvalue problem}) and at the EPD is different
from any other regular frequency in the dispersion diagram since the
system matrix contains repeated eigenvalues associated with one eigenvector.
Thus, the time-domain response of the circuit at the second-order
EPD is expected to be in the form of $\boldsymbol{\Psi}(t)\propto te^{j\omega_{\mathit{e}}t}$,
as it is indeed shown in Fig. \ref{Fig: EPD-L-Neg}(c). The voltage
grows linearly by increasing time, whereas the oscillation frequency
is constant. This remarkable feature is peculiar to an EPD, and it
is the result of coalescing eigenvalues and eigenvectors that also
corresponds to a double pole in the circuit (or zero, depending on
what is observed). We take a fast Fourier transform (FFT) of the voltage
$v_{1}(t)$ to show the frequency spectrum, and the calculated result
is illustrated in Fig. \ref{Fig: EPD-L-Neg}(d). The result is calculated
from $40\mathrm{\:kHz}$ to $120\:\mathrm{kHz}$ by applying an FFT
with $10^{6}$ samples in the time window of $0\:\mathrm{ms}$ to
$0.4\:\mathrm{ms}$. The numerically observed oscillation angular
frequency is $f_{0}=\omega_{o}/(2\pi)=81.63\mathrm{\:kHz}$, that
shows the frequency corresponds to the maximum value in \ref{Fig: EPD-L-Neg}(d).
The numerically obtained value is in good agreement with the theoretical
value calculated above.

So far, we have used the gyrator-based circuit to measure the perturbation
near EPD by varying the gyrator resistance. Next, we analyze the circuit's
sensitivity to independent perturbations in the positive values of
\textit{both} capacitances. We change the capacitance value on each
resonator independently and calculate the eigenfrequencies' real and
imaginary parts. The three-dimensional result for the calculated eigenfrequencies
is illustrated in Figs. \ref{Fig: EPD-L-Neg}(e), and (f). The elevation
value of any point on the surface shows the eigenfrequency, and the
associated color helps to recognize it conveniently. In these figures,
only the two solutions with $\mathrm{Re}(\omega)>0$ are illustrated.
Although the resonance frequency of each resonator in this paper is
imaginary, in the specific range of $C_{1}$ and $C_{2}$, the EPD
frequency is purely real. To utilize these calculated results, the
flat version of the three-dimensional diagram for the real part is
provided in Figs. \ref{Fig: EPD-L-Neg}(g), and (h) for higher and
lower eigenfrequency. These figures can help designers in the design
procedure to select the proper value for components to achieve the
desired real resonance frequency. The intersection of two surfaces
(eigenfrequencies surface and surface of constant $z$ plane) is a
one-dimensional curve. Therefore, there is a different set of values
for capacitances to make oscillation at a certain frequency. Moreover,
the intersection of the higher eigenfrequencies surface and lower
eigenfrequencies surface indicates the possible EPD that various combinations
of capacitances values can yield. Designers can use these figures
and pick a proper value in the design steps according to their practical
limitations.

\subsection{Negative Capacitances $C_{1}$ and $C_{2}$\label{subsec:Negative-Capacitances}}

\begin{figure*}[tbh]
\centering{}\includegraphics[width=7in]{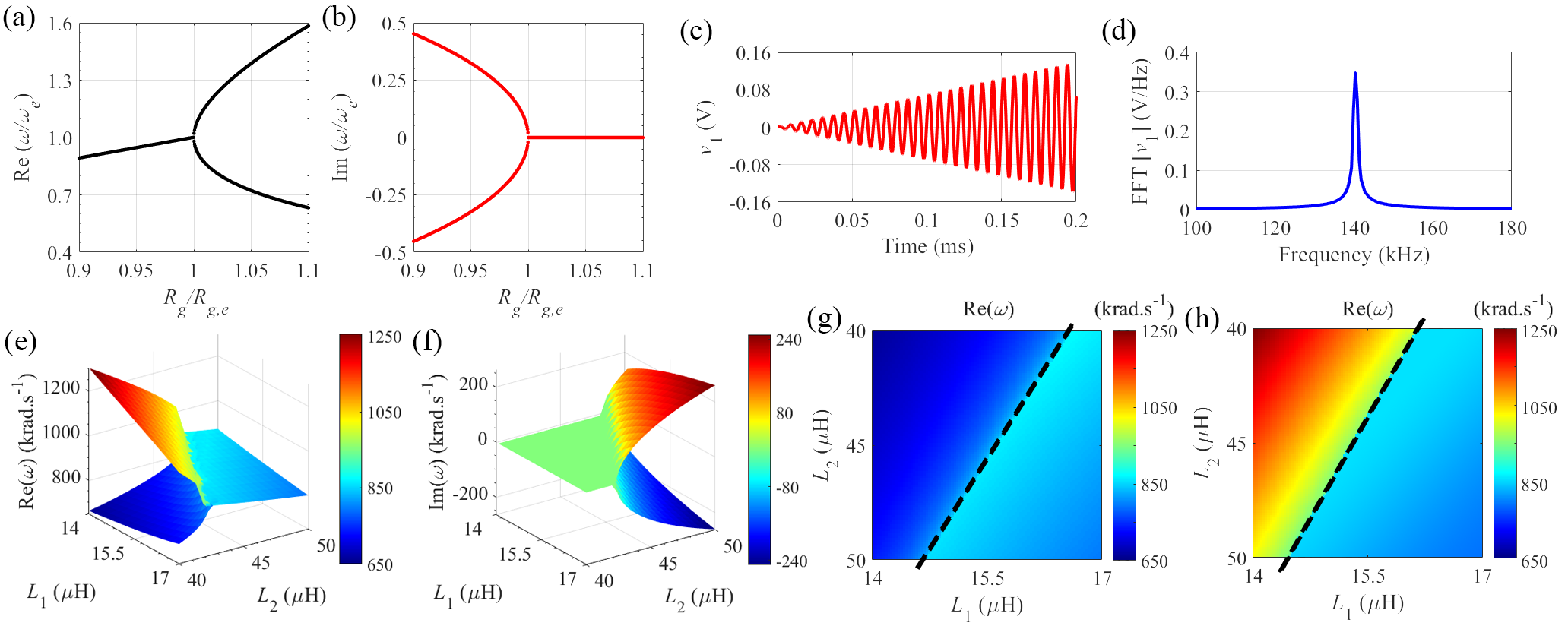}\caption{The sensitivity of the (a) real and (b) imaginary parts of the eigenfrequencies
to gyration resistance perturbation. Voltage $v_{1}(t)$ under the
EPD condition in the (c) time-domain, and (d) frequency-domain. The
frequency-domain result is calculated from $100\mathrm{\:kHz}$ to
$180\:\mathrm{kHz}$ by applying an FFT with $10^{6}$ samples in
the time window of $0\:\mathrm{ms}$ to $0.2\:\mathrm{ms}$. The three-dimensional
plot of the (e) real and (f) imaginary parts of the eigenfrequencies
to $L_{1}$ and $L_{2}$ perturbation. The real part of eigenfrequencies
for (g) higher and (h) lower value of resonance frequencies which
colormap show the resonance frequency value. The black dashed line
in these plots shows the EPD.\label{Fig: EPD-C-Neg}}
\end{figure*}
In the following section, we consider another condition in which negative
capacitances are used on both resonators; so $\omega_{g}^{2}>0$.
Using the mentioned presumption, the first and second terms in Eq.
(\ref{eq: EPD_a}) are negative because of the imaginary value of
resonance frequencies of resonators, and the third term is positive.
So, if the EPD condition is met, the sign of $a$ in Eq. (\ref{eq: EPD_a})
indicates whether the eigenfrequency is real or imaginary. According
to Eq. (\ref{eq: Eigenfrequencies}), if $\left|\omega_{01}^{2}+\omega_{02}^{2}\right|<\omega_{g}^{2}$
we get a real value for the EPD frequency, and if $\left|\omega_{01}^{2}+\omega_{02}^{2}\right|>\omega_{g}^{2}$,
the EPD frequency is imaginary.

Different combinations of values for the circuit's components can
satisfy the EPD condition demonstrated in Eq. (\ref{eq: EPDCondition}),
and here as an example, we use this set of values: $C_{1}=-47\mathrm{\:nF}$,
$C_{2}=-47\mathrm{\:nF},$ $L_{2}=47\mathrm{\:\mu H}$, and $R_{g}=50\:\Omega$.
The inductance value on the left resonator is calculated by solving
the resulting quadratic equation from Eq. (\ref{eq: EPDCondition}).
In the presented example, $L_{1}$ can be a sensing inductor in a
system. According to Eq. (\ref{eq: EPDCondition}), two different
values for inductance in the first resonator are calculated after
solving the quadratic equation. We consider $L_{1,\mathit{e}}=15.87\mathrm{\:\mu H}$
for this example, so both $\omega_{01}^{2}$ and $\omega_{02}^{2}$
have negative values, with $\omega_{01}=-j1.16\:\mathrm{Mrad/s}$,
and $\omega_{02}=-j672.82\:\mathrm{krad/s}$. Then, we obtain a positive
value for $a$ in Eq. (\ref{eq: EPD_a}), leading to a real EPD angular
frequency of $\omega_{\mathit{e}}=881.6\mathrm{\:krad/s}$. The results
in Figs. \ref{Fig: EPD-C-Neg}(a), and (b) shows the two branches
of the real and imaginary parts of eigenfrequencies obtained by perturbing
$R_{g}$ near the value that made the EPD.

The time-domain simulation result by using the Keysight ADS with an
initial condition $v_{1}(0)=1\:\mathrm{mV}$ is presented in Fig.
\ref{Fig: EPD-C-Neg}(c). The voltage $v_{1}(t)$ is calculated in
the time interval of $0$ ms to $0.2$ ms. Fig. \ref{Fig: EPD-C-Neg}(c)
shows $v_{1}(t)$ is growing linearly by increasing time. The growing
signal demonstrates the circuit eigenvalues coalesce, and the output
rises linearly at the second-order EPD frequency. In order to evaluate
the oscillation frequency from the time-domain simulation, we take
an FFT of voltage $v_{1}(t)$ from $100\mathrm{\:kHz}$ to $180\:\mathrm{kHz}$
using $10^{6}$ samples in the time window of $0\:\mathrm{ms}$ to
$0.2\:\mathrm{ms}$. The calculated spectrum is shown in Fig. \ref{Fig: EPD-C-Neg}(d),
showing an oscillation frequency of $f_{0}=\omega_{o}/(2\pi)=140.31\mathrm{\:kHz}$,
which is in good agreement with the calculated theoretical value obtained
from Eq. (\ref{eq: EPDFrequency}).

In the following step, we investigate the circuit's sensitivity to
independent perturbation in the value of both inductances. The real
and imaginary parts of eigenfrequencies are calculated when the values
of the inductances are changed. The three-dimensional eigenfrequencies
map of the two solutions with $\mathrm{Re}(\omega)>0$ is shown in
Figs. \ref{Fig: EPD-C-Neg}(e), and (f). In order to provide a better
representation, the flat view of the three-dimensional diagram for
the real part is shown in Figs. \ref{Fig: EPD-C-Neg}(g), and (h)
for higher and lower eigenfrequencies.

\subsection{Negative Inductance on One Side and Negative Capacitance on The Other
Side}

In this last case, different constraints for components value are
considered. We assume a component with a negative value on one side
(capacitance/inductance) and the other component with a negative value
on the other side (inductance/capacitance); hence, in this case $\omega_{g}^{2}<0$.
For instance, we consider a negative inductance on the right resonator
and a negative capacitance on the left resonator. It is evident that
in this case, we have two unstable resonators when they are uncoupled.
When two resonators are coupled, EPD should satisfy Eq. (\ref{eq: EPD_b}).
According to Eq. (\ref{eq: EPDFrequency}), all terms inside the square
root are negative, and the sum of negative values is always negative.
As a result, it is impossible to achieve an EPD with a real eigenfrequency
under the assumption mentioned above. Since we focus on cases with
real EPD frequency in this paper, we will skip considering this condition
in the rest of the paper.

\section{Frequency-Domain Analysis of The Resonances in Lossless Gyrator-Based
Circuit}

\begin{figure}[t]
\centering{}\includegraphics[width=3.5in]{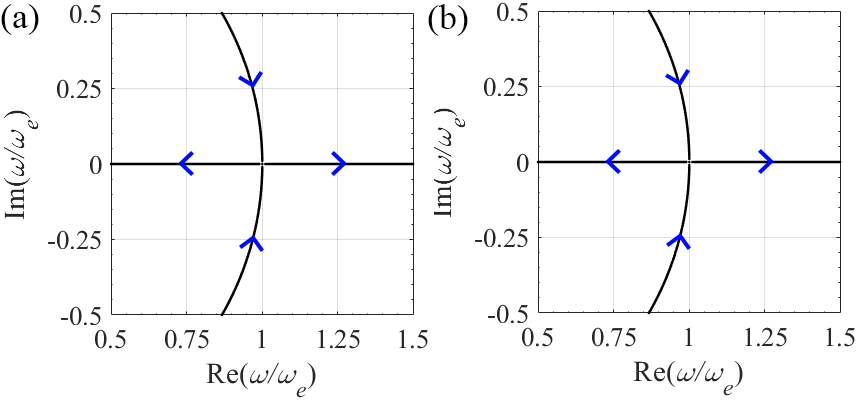}\caption{Root locus of zeros of $Z_{total}(\omega)=0$ showing the real and
imaginary parts of resonance frequencies of the circuit when varying
gyration resistance (arrows represent growing $R_{g}$ values). In
these figures, we consider two cases, with a negative value of (a)
both inductances and (b) both capacitances, discussed in Section \ref{sec:Lossless-EPD-Condition}.
At the EPD, the system's total impedance is $Z_{total}(\omega)\propto(\omega-\omega_{\mathit{e}})^{2}$;
hence it exhibits a double zero at $\omega_{\mathit{e}}$.\label{Fig: RootLucas}}
\end{figure}

We demonstrate how the EPD regime is associated with a special kind
of circuit's resonance, directly observed in frequency-domain circuit
analysis. First, we calculate the transferred impedance on the left
port of the gyrator in Fig. \ref{Fig: Circuit}, which is

\begin{equation}
Z_{trans}(\omega)=\frac{R_{g}^{2}}{Z_{2}(\omega)},\label{eq:Ztrans}
\end{equation}
where $Z_{2}(\omega)=j\omega L_{2}+1/(j\omega C_{2})$ is the impedance
of LC tank on the right side of the gyrator. The total impedance observed
from the input port (see Fig. \ref{Fig: Circuit}) is

\begin{equation}
Z_{total}(\omega)\triangleq Z_{1}(\omega)+Z_{trans}(\omega)=Z_{1}(\omega)+\frac{R_{g}^{2}}{Z_{2}(\omega)},\label{eq: Ztotal}
\end{equation}
where $Z_{1}(\omega)=j\omega L_{1}+1/(j\omega C_{1})$ is the impedance
of LC tank on the left side of the gyrator. The complex-valued resonance
frequencies of the circuit are calculated by imposing $Z_{total}(\omega)=0$.
Fig. \ref{Fig: RootLucas} shows the zeros of such total impedance
$Z_{total}(\omega)$ for various gyration resistance values (arrows
represent growing $R_{g}$ values). When considering the EPD gyrator
resistance $R_{g}=R_{g,\mathit{e}}=50\,\Omega$, one has $Z_{total}(\omega)\propto(\omega-\omega_{\mathit{e}})^{2}$,
i.e., the two zeros coincide with the EPD angular frequency $\omega_{\mathrm{\mathit{e}}}$,
that is also the point where the two curves in Fig. \ref{Fig: RootLucas}
meet. For gyrator resistances $R_{g}<R_{g,\mathit{e}}$, the two resonance
angular frequencies are complex conjugate, consistent with the result
in Fig. \ref{Fig: RootLucas}. Also, for gyrator resistances such
that $R_{g}>R_{g,\mathit{e}}$, the two resonance angular frequencies
are purely real. In other words, the EPD frequency coincides with
double zeros of the frequency spectrum, or double poles, depending
on the way the circuit is described.

\section{EPD in The Lossy Gyrator-Based Circuit\label{sec: LossyCircuit}}

\begin{figure}[t]
\centering{}\includegraphics[width=3.5in]{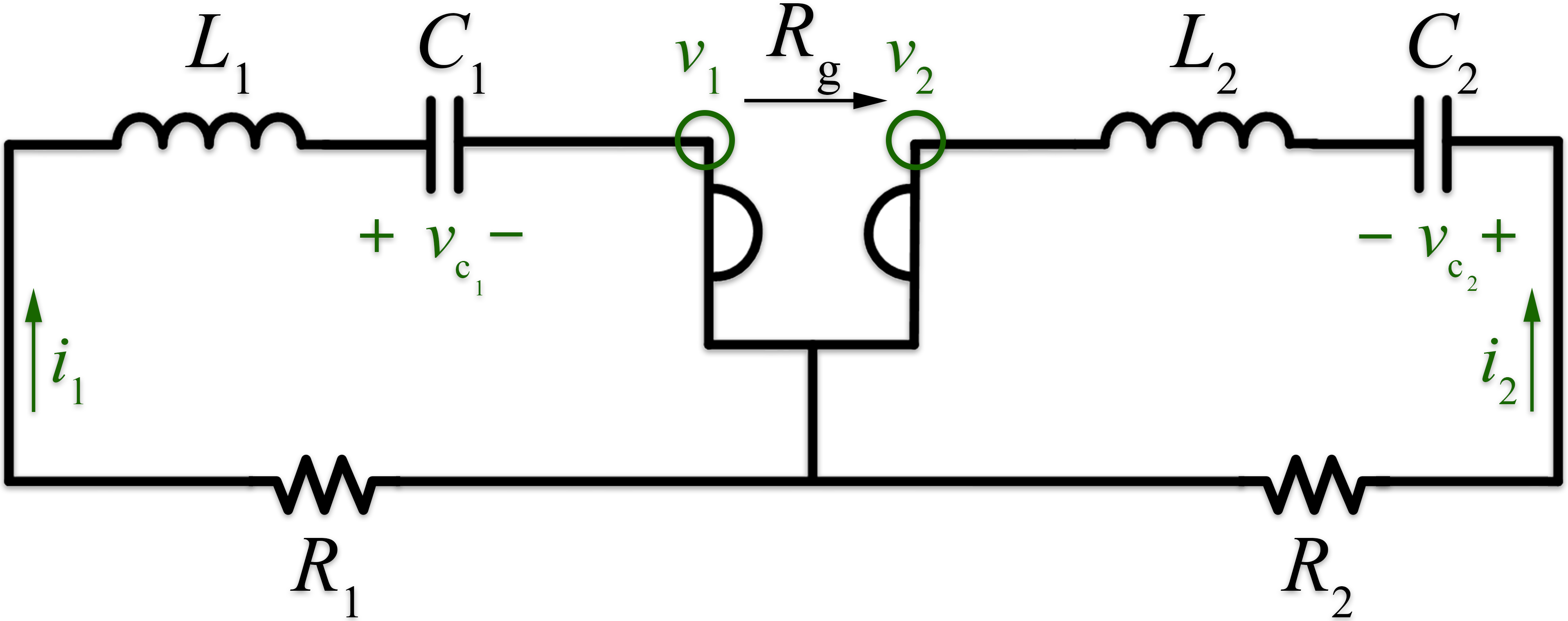}\caption{Schematic view of the lossy gyrator-based circuit, with a resistor
in each resonator.\label{Fig: LossyCircuit}}
\end{figure}

The following section analyzes the EPD condition in the gyrator-based
circuit by accounting for series resistors $R_{1}$ and $R_{2}$ in
resonators as illustrated in Fig. \ref{Fig: LossyCircuit}. A procedure
analogous to the one discussed earlier, using the same state vector
$\boldsymbol{\Psi}\equiv[Q_{1},Q_{2},\dot{Q}_{1},\dot{Q}_{2}]^{T}$,
leads to \cite{nikzamir2021demonstration}

\begin{equation}
\mathcal{\mathrm{\frac{d\boldsymbol{\Psi}}{dt}=\mathbf{\underline{M}}\boldsymbol{\Psi},\;\;}}\mathbf{\underline{M}}=\left(\begin{array}{cccc}
0 & 0 & 1 & 0\\
0 & 0 & 0 & 1\\
-\omega_{01}^{2} & 0 & -\gamma_{1} & \frac{R_{g}}{L_{1}}\\
0 & -\omega_{02}^{2} & -\frac{R_{g}}{L_{2}} & -\gamma_{2}
\end{array}\right).
\end{equation}

In the presented lossy circuit matrix, $\gamma_{1}=R_{1}/L_{1}$,
and $\gamma_{2}=R_{2}/L_{2}$ determine losses in each resonator.
These eigenfrequencies of the circuit are found by solving the below
characteristic equation,

\begin{equation}
\begin{array}{c}
\omega^{4}-j\omega^{3}\left(\gamma_{1}-\gamma_{2}\right)-\omega^{2}\left(\omega_{01}^{2}+\omega_{02}^{2}+\gamma_{1}\gamma_{2}+\frac{R_{g}^{2}}{L_{1}L_{2}}\right)\\
+j\omega\left(\gamma_{1}\omega_{02}^{2}+\gamma_{2}\omega_{01}^{2}\right)+\omega_{01}^{2}\omega_{02}^{2}=0.
\end{array}\label{eq:LossyCharacteristic}
\end{equation}

The coefficients of the odd-power terms of the angular eigenfrequency
in the characteristic equation are imaginary; therefore, $\omega$
and $-\omega^{*}$ are both roots of the characteristic equation.
In order to obtain a stable circuit with real-valued eigenfrequencies,
the coefficients of the odd-power terms in the characteristic equation
Eq. (\ref{eq:LossyCharacteristic}), $-j(\gamma_{1}-\gamma_{2})$
and $j(\gamma_{1}\omega_{02}^{2}+\gamma_{2}\omega_{01}^{2})$, should
vanish, otherwise a complex eigenfrequency is needed to satisfy the
characteristic equation. The coefficient of the $\omega^{3}$ term
is zero when $\gamma_{1}=\gamma_{2}$, but according to this condition,
the coefficient of the $\omega$ term is non-zero because $\omega_{01}^{2}$
and $\omega_{02}^{2}$ are both negative. Moreover, the coefficient
of the $\omega$ term never vanishes when both resonators are lossy
because both $\omega_{01}^{2}$ and $\omega_{02}^{2}$ have the same
sign. Consequently, it is not possible to have all real-valued coefficients
in the characteristic polynomials, except when $\gamma_{1}=\gamma_{2}=0$,
which corresponds to a lossless circuit.

\subsection{RLC Resonators With Negative Inductances $L_{1}$ and $L_{2}$}

\begin{figure}[t]
	\centering{}\includegraphics[width=3.5in]{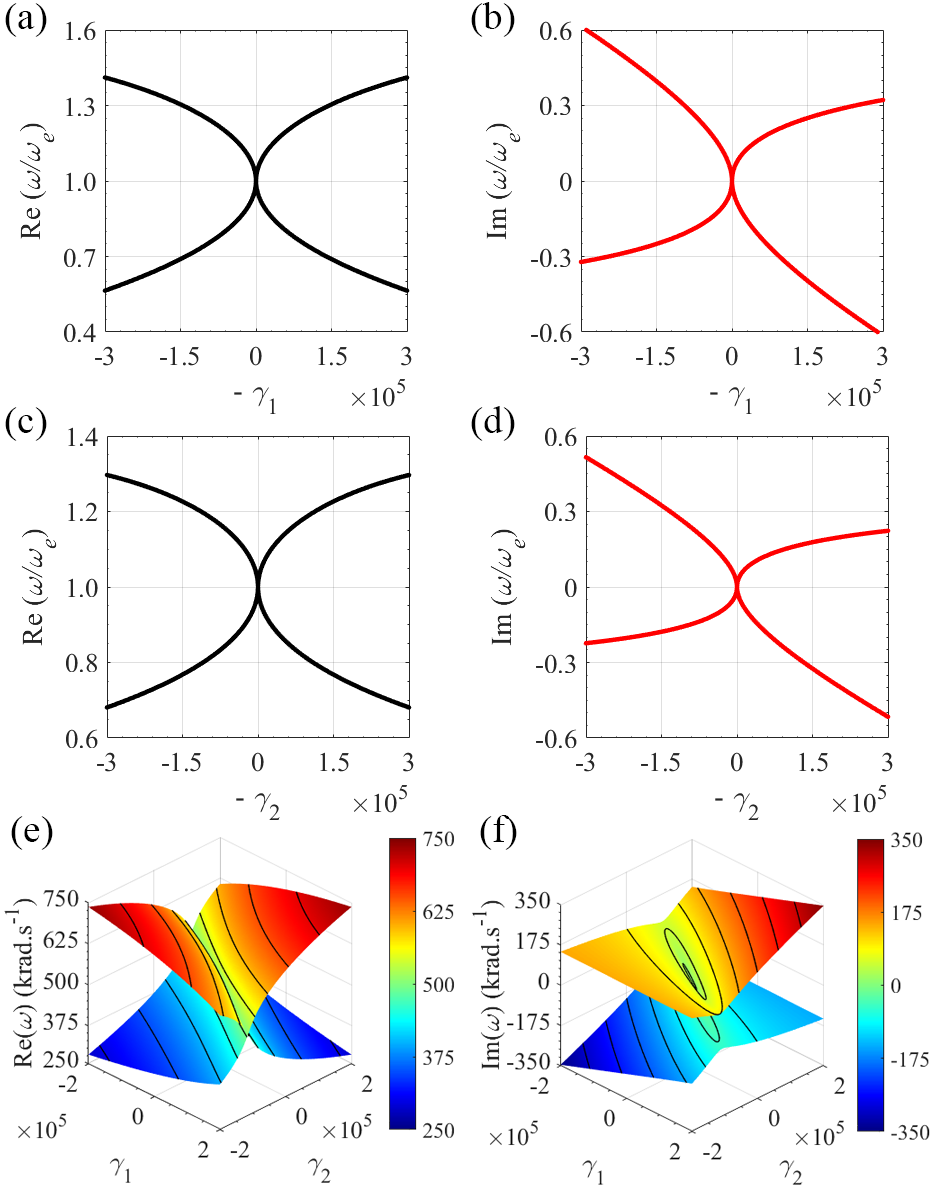}\caption{Case with negative value of the inductances on both resonators. Variation
		of (a) real and (b) imaginary parts of the angular eigenfrequencies
		to a resistor perturbation on the left resonator, i.e., when $-\gamma_{1}$
		changes and $\gamma_{2}=0$. (c) and (d), as in (a) and (b), but the
		resistor perturbation is on the right resonator, i.e., $-\gamma_{2}$
		changes and $\gamma_{1}=0$. Variation of (e) real and (f) imaginary
		parts of the angular eigenfrequencies to independent resistor perturbation
		on the both sides.\label{Fig: Lossy-L-Neg}}
\end{figure}

In the first case, we assume inductances with negative values. In
Figs. \ref{Fig: Lossy-L-Neg}(a) and (b), $\gamma_{1}$ is perturbed
while we assume $\gamma_{2}=0$, whereas in Figs. \ref{Fig: Lossy-L-Neg}(c),
and (d), $\gamma_{2}$ is perturbed while $\gamma_{1}=0$. These four
figures present the real and imaginary parts of eigenfrequencies when
the resistances $R_{1}$ and $R_{2}$ are perturbed individually.
We use the same values for the circuit components as already used
in the lossless circuit presented in Subsection \ref{subsec:Negative-Inductances}.
The normalization term $\omega_{\mathit{e}}$ is the EPD angular frequency
obtained when $\gamma_{1}=\gamma_{2}=0$, which is the same EPD frequency
of the lossless circuit. In this case, losses in the circuit are represented
by negative $\gamma_{1}$ and $\gamma_{2}$ since $L_{1}$, and $L_{2}$
are negative, so the right half side of the figure axes show the loss
and the left half side of the axes represent the gain in the circuit
through a negative resistance. In Figs. \ref{Fig: Lossy-L-Neg}(a)-(d),
we recognize the bifurcations of the real and imaginary parts of the
eigenfrequencies, so the circuit is extremely sensitive to variations
of resistances in the vicinity of EPD. By perturbing $\gamma_{1}$
or $\gamma_{2}$ away from $\gamma_{1}=\gamma_{2}=0$, the circuit
becomes unstable, and it begins to self oscillate at a frequency associated
with the real part of the unstable angular eigenfrequency. In addition,
we show the real and imaginary parts of the eigenfrequencies by separately
perturbing the resistances in both sides in \ref{Fig: Lossy-L-Neg}(e)-(f).
The black contour lines in these three-dimensional figures show constant
real and imaginary parts of the eigenfrequencies. We observe that
by adding either loss or gain, the circuit becomes unstable. Instability
in the circuit is not due to the instability of the uncoupled resonators,
but rather it is unstable because of the addition of losses, as was
the case in \cite{nikzamir2021demonstration} for different configurations.
When $\gamma_{1}$ or $\gamma_{2}$ is perturbed from the EPD, the
oscillation frequency is shifted from the EPD frequency, and it could
be measured for sensing applications. The eigenfrequency with a negative
imaginary part is associated with an exponentially growing signal
(instability). Considering the existence of instability, there are
a few possible ways of operations: preventing the system from reaching
saturation by switching off the circuit, partially compensating for
losses, or make the circuit an oscillator. In the partial compensation
scheme, the instability effect due to losses in the circuit can be
counterbalanced by adding an independent series gain to each resonator.
A negative resistance can be easily implemented using the same opamp-based
circuit designed to achieve negative inductance and capacitance. This
issue is out of the scope of this paper, and it seems a complicated
strategy for stability. We believe that exploiting the system's instability
may be an excellent strategy to design sensitive oscillators that
work as sensors; this could be the subject of future investigations.

\begin{figure}[t]
	\centering{}\includegraphics[width=3.5in]{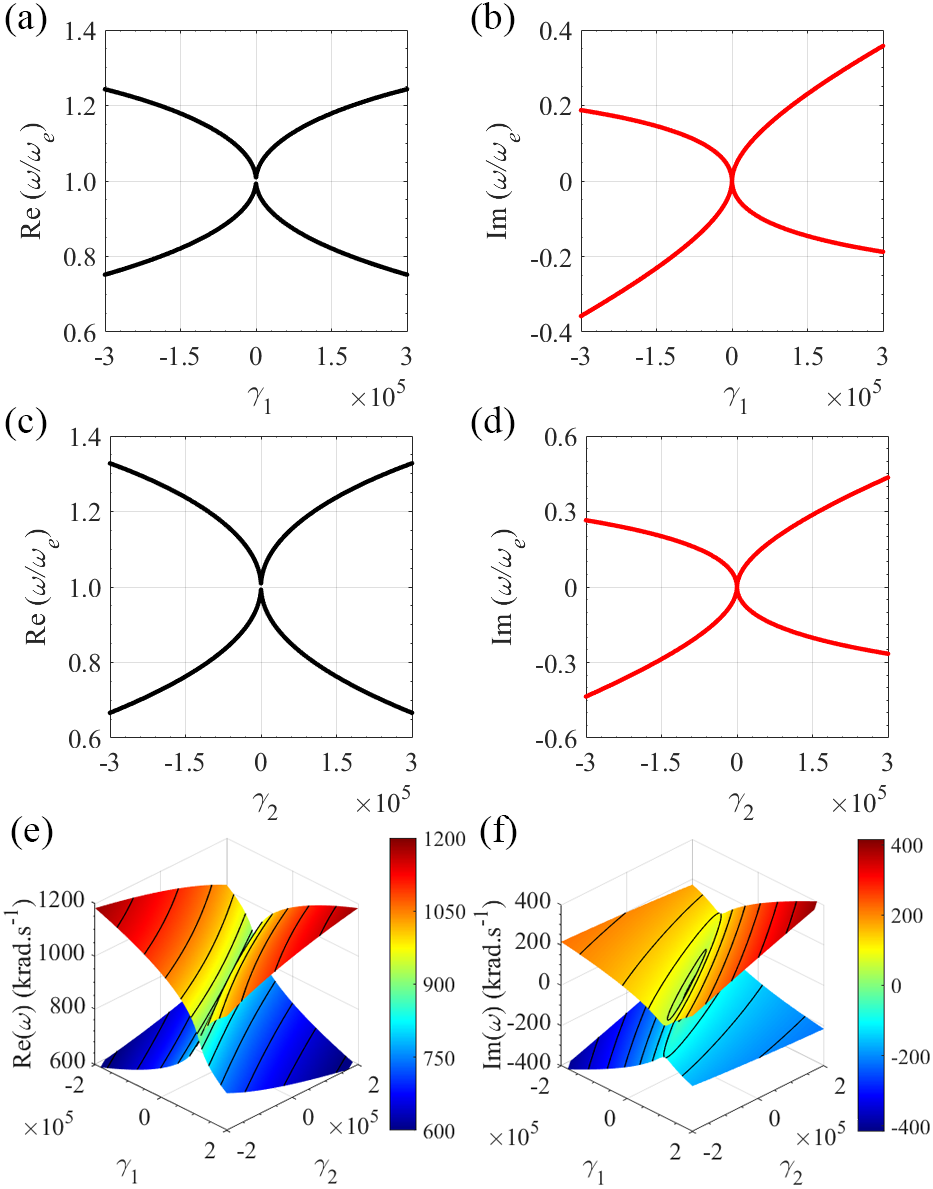}\caption{As in Fig. \ref{Fig: Lossy-L-Neg}, except that these results pertain
		to the case where both capacitances are negative.\label{Fig: Lossy-C-Neg}}
\end{figure}

\subsection{RLC Resonators With Negative Capacitances $C_{1}$ and $C_{2}$}

In the second case, we consider the negative value for capacitances.
In Figs. \ref{Fig: Lossy-C-Neg}(a) and (b), $\gamma_{1}$ is perturbed
while we consider $\gamma_{2}=0$ and in Figs. \ref{Fig: Lossy-C-Neg}(c),
and (d), $\gamma_{2}$ is perturbed while $\gamma_{1}=0$. These figures
show the real and imaginary parts of eigenfrequencies when each resistor
is perturbed individually. We use the same values for the circuit
components as used earlier in the lossless circuit shown in Subsection
\ref{subsec:Negative-Capacitances}, and the EPD angular frequency
is obtained for these circuit parameters when $\gamma_{1}=\gamma_{2}=0$,
which is the same EPD frequency of the lossless circuit. In Figs.
\ref{Fig: Lossy-C-Neg}(a)-(d), we observe the bifurcations of the
real and imaginary parts of the eigenfrequencies, so the circuit is
exhibits extreme sensitivity to resistance value variations in the
vicinity of EPD. We show the real and imaginary parts of the eigenfrequencies
by independently changing the resistances in both sides in \ref{Fig: Lossy-C-Neg}(e)-(f).
The black contour lines in these three-dimensional figures show constant
real and imaginary parts of the eigenfrequencies. Angular eigenfrequencies
are complex-valued when perturbing $\gamma_{1}$ and $\gamma_{2}$
away from $\gamma_{1}=\gamma_{2}=0$; hence the circuit gets unstable
and it starts to oscillate at a fundamental frequency associated with
the real part of the unstable angular eigenfrequency. In Figs. \ref{Fig: Lossy-C-Neg}(a)-(f),
both conditions $\gamma_{1}>0$ and $\gamma_{2}>0$ represent loss,
whereas the conditions $\gamma_{1}<0$ and $\gamma_{2}<0$ represent
gain in the circuit through a negative resistance.

\begin{figure*}[t]
	\centering{}\includegraphics[width=7in]{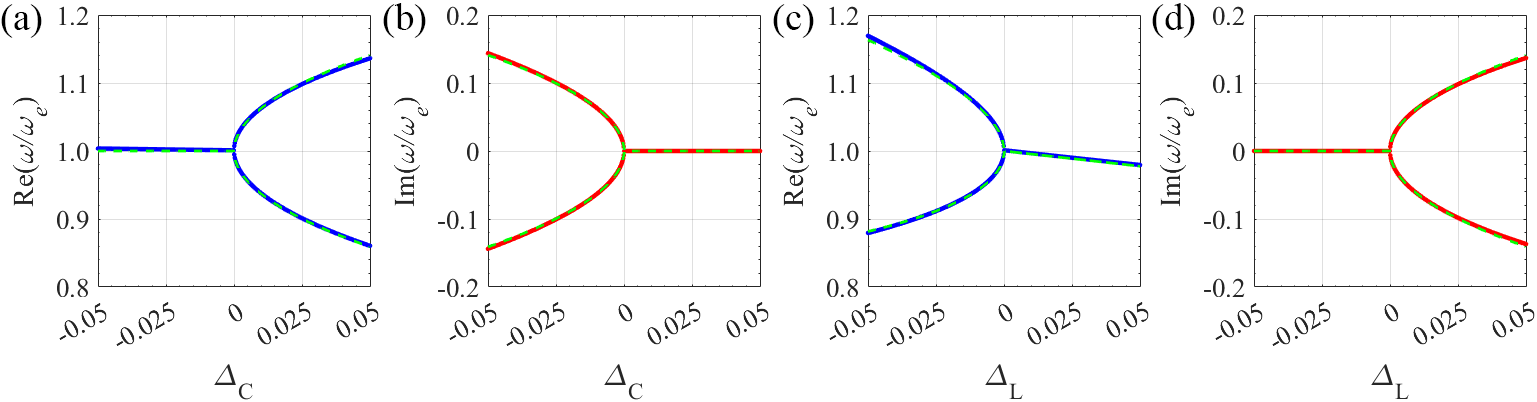}\caption{Sensitivity of (a) real and (b) imaginary parts of the eigenfrequencies
		to a capacitance perturbation (solid lines), $\Delta_{\mathrm{C}}=(C_{\mathrm{1}}-C_{\mathrm{1,\mathit{e}}})/C_{\mathrm{1,\mathit{e}}}$,
		while the inductances values on both sides are negative. Dashed lines
		show the perturbed eigenfrequencies according to the Puiseux expansion
		up to its first order. Sensitivity of (c) real and (d) imaginary parts
		of the eigenfrequencies to an inductance perturbation (solid lines),
		$\Delta_{\mathrm{L}}=(L_{\mathrm{1}}-L_{\mathrm{1,\mathit{e}}})/L_{\mathrm{1,\mathit{e}}}$,
		while the capacitances values on both sides are negative. Dashed lines
		show the perturbed eigenfrequencies according to Puiseux expansion
		up to its second order.\label{Fig: Puiseux}}
\end{figure*}

\section{High-Sensitivity and Puiseux Fractional Power Series Expansion}

Eigenfrequencies at EPDs are extremely sensitive to perturbations
of the circuit elements, a property that is peculiar to the EPD condition.
We study the circuit under EPD perturbation to investigate the circuit's
sensitivity near the EPD. We will demonstrate how small perturbations
in a component's value perturb the eigenfrequencies of the circuit.
In order to do this analysis, the relative circuit perturbation $\Delta_{\mathrm{X}}$
is defined as

\begin{equation}
\Delta_{\mathrm{X}}=\frac{X-X_{\mathit{e}}}{X_{\mathit{e}}},\label{eq: perturbation_touchstone}
\end{equation}
where $X$ is the perturbed parameter value, and $X_{\mathit{e}}$
is its unperturbed value that provides the EPD. The perturbation in
$\Delta_{\mathrm{X}}$ value leads to a perturbed circuit matrix $\mathbf{\underline{M}}(\Delta_{\mathrm{X}})$.
We demonstrate the extreme sensitivity to extrinsic perturbation by
resorting to the general theory of EPD and utilizing the Puiseux fractional
power series expansion \cite{Kato1995Perturbation}. Accordingly,
when a small relative perturbation in component value $\Delta_{\mathrm{X}}$
is applied, the resulting two different eigenfrequencies $\omega_{\mathrm{\mathit{p}}}(\Delta_{\mathrm{X}})$,
with $p=1,2$ are estimated using the convergent Puiseux series. Here
we provide the first two terms to estimate the eigenfrequencies near
an EPD, using the explicit formulas given in \cite{Welters2011OnExplicit},

\begin{equation}
\omega_{p}(\Delta_{\mathrm{X}})\approx\omega_{\mathit{e}}+(-1)^{p}\alpha_{1}\sqrt{\Delta_{\mathrm{X}}}+\alpha_{2}\Delta_{\mathrm{X}},\label{eq: Puiseux}
\end{equation}

\begin{equation}
\alpha_{1}=\left.\sqrt{-\frac{\frac{\partial H(\Delta_{\mathrm{X}},\omega)}{\partial\Delta_{\mathrm{X}}}}{\frac{1}{2!}\frac{\partial^{2}H(\Delta_{\mathrm{X}},\omega)}{\partial\omega^{2}}}}\right|_{\Delta_{\mathrm{X}}=0,\,\omega=\omega_{\mathrm{\mathit{e}}}},\label{eq: PuiseuxCoeff1}
\end{equation}

\begin{equation}
\alpha_{2}=\left.-\frac{\alpha_{1}^{2}\frac{1}{3!}\frac{\partial^{3}H(\Delta_{\mathrm{X}},\omega)}{\partial\omega^{3}}+\frac{\partial^{2}H(\Delta_{\mathrm{X}},\omega)}{\partial\omega\partial\Delta_{\mathrm{X}}}}{\frac{\partial^{2}H(\Delta_{\mathrm{X}},\omega)}{\partial\omega^{2}}}\right|_{\Delta_{\mathrm{X}}=0,\,\omega=\omega_{\mathrm{\mathit{e}}}},\label{eq: PuiseuxCoeff2}
\end{equation}
where $H(\Delta_{\mathrm{X}},\omega)=\mathrm{det}[\mathbf{\underline{M}}(\Delta_{\mathrm{X}})-j\omega\underline{\mathbf{I}}]$,
and $\alpha_{1}$, and $\alpha_{2}$ are first- and second-order coefficients
respectively. Eq. (\ref{eq: Puiseux}) indicates that for a tiny perturbation
in component value $\Delta_{\mathrm{X}}\ll1$ the eigenvalues change
sharply from their original degenerate value due to the square root
function, which is an essential characteristic of second-order EPD.

Typically, the inductor or capacitor changes in response to an external
perturbation of the parameter of interest, leading to a shift in resonance
frequency. We consider variations of $L_{\mathrm{1}}$, or $C_{\mathrm{1}}$,
one at the time, and the calculated real and imaginary parts of the
eigenfrequencies near the EPD is shown in Figs. \ref{Fig: Puiseux}.
In the first case, the perturbation parameter is the capacitance,
$\Delta_{\mathrm{C}}=(C_{\mathrm{1}}-C_{\mathrm{1,\mathit{e}}})/C_{\mathrm{1,\mathit{e}}}$,
and a negative value for both inductances is assumed, so the first-order
Puiseux expansion coefficient is calculated as $\alpha_{1}=3.228\times10^{5}\:\mathrm{rad/s}$.
To calculate the coefficients, we used the components value utilized
in Subsection \ref{subsec:Negative-Inductances}. Figs. \ref{Fig: Puiseux}(a)
and (b) exhibit the real and imaginary parts of the perturbed eigenfrequencies
$\omega$ obtained from the eigenvalue problem after perturbing $\Delta_{\mathrm{C}}$.
Furthermore, green dashed lines in these figures demonstrate such
perturbed eigenfrequencies are well estimated with high accuracy by
using the Puiseux expansion truncated at its first order. For a negative
but small value of $\Delta_{\mathrm{C}}$, the imaginary part of the
eigenfrequencies experiences a rapid change, and its real part remains
constant. On the other hand, a very small positive value of $\Delta_{\mathrm{C}}$
causes a sharp change in the real part of the eigenfrequencies while
its imaginary part remains unchanged.

In the second example, the inductance value in the left resonator
is considered as a perturbed parameter, $\Delta_{\mathrm{L}}=(L_{\mathrm{1}}-L_{\mathrm{1,\mathit{e}}})/L_{\mathrm{1,\mathit{e}}}$,
whereas capacitances values are both negative. By using Eqs. (\ref{eq: PuiseuxCoeff1}),
and (\ref{eq: PuiseuxCoeff2}) and using the components value utilized
in Subsection \ref{subsec:Negative-Capacitances}, the coefficients
of the Puiseux expansion are calculated as $\alpha_{1}=j5.548\times10^{5}\:\mathrm{rad/s}$
and $\alpha_{2}=-3.960\times10^{5}\:\mathrm{rad/s}$. The calculated
results in Figs. \ref{Fig: Puiseux}(c), and (d) show the two branches
(solid lines) of the exact perturbed eigenfrequencies evaluated from
the eigenvalue problem when the external perturbation is applied to
the circuit. This figure shows that the perturbed eigenfrequencies
are estimated accurately by applying the Puiseux expansion truncated
at its second order (dashed lines). For a tiny value of positive perturbation,
the imaginary part of the eigenfrequencies undergoes sharp changes,
while its real part remains approximately unchanged. However, a small
negative perturbation in the inductance value rapidly changes the
real part of the two eigenfrequencies away from the EPD eigenfrequency.
The bifurcation in the diagram, described by a square root, is the
most exceptional physical property associated with the EPD. It can
be employed to devise ultra-sensitive sensors for various applications
\cite{Hodaei2017Enhanced,sakhdari2018ultrasensitive,Rouhi2020Exceptional,li2021enhancing}.

\section{Sensing Scenario for Liquid Content Measurement}

In recent years various well-established techniques has been proposed
to measure the liquid level such as light-reflection sensors \cite{azzam1980light},
chirped fiber Bragg grating \cite{yun2007highly,vorathin2018novel},
fiber optic sensors \cite{khaliq2001fiber,golnabi2004design,lin2014low},
ultrasonic Lamb waves \cite{sakharov2003liquid}, and capacitive sensors
\cite{toth1997planar,canbolat2009novel,chetpattananondh2014self,hanni2020novel}.
The use of a capacitive sensor is a well-known method for liquid level
measurement \cite{kumar2014review}. This kind of sensor has been
proven to be stable, can be assembled using various materials, and
can provide high resolution \cite{loizou2016water}. The principle
of operation of capacitive sensors is that it converts a variation
in position, or material characteristics, into measurable electrical
signals \cite{bera2006low}. Capacitive sensors are operated by changing
any of the three main parameters: relative dielectric constant, area
of capacitive plates, and distance between the plates. In conventional
methods, a capacitive liquid level detector can sense the fluid level
by measuring variations in capacitance made between two conducting
plates embedded outside a non-conducting tank or immersed in the liquid
\cite{terzic2012neural,kumar2014review}. The same concept applies
when the liquid occupies a varying volume percentage as a mixture's
component.

\begin{figure}[!t]
\centering{}\includegraphics[width=3.5in]{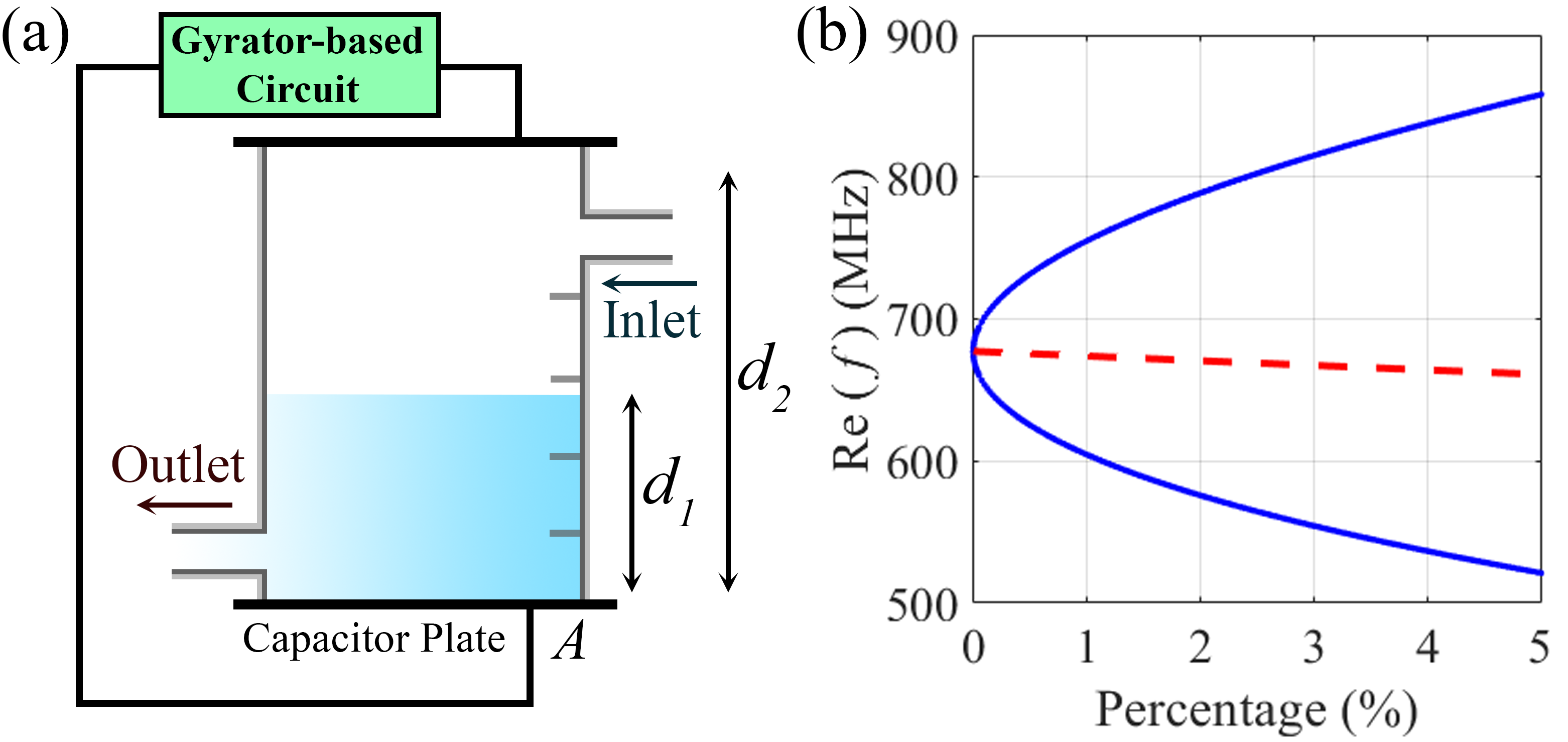}\caption{(a) Schematic illustration of a device for liquid level measurement.
(b) The EPD is designed at a given level of water content ($0\ensuremath{\%}
$ in the figure). The solid blue line in the plot shows the two resonance
frequencies of the gyrator-based circuit versus water level variation
with very high sensitivity near $0\ensuremath{\%}
$. Also, the red dashed line shows the resonance frequency of a single
resonator when the water content changes. The EPD-based circuit and
the single LC resonator have the same resonance frequency at $0\ensuremath{\%}
$. It is clear that the EPD-based circuit provides much higher sensitivity
to the capacitance perturbation than the single LC resonator.\label{Fig: Sensor}}
\end{figure}

The following design will provide the schematic of a practical scenario
to investigate the gyrator-based circuit application for physical
parameter measurement. We provide the required setup and the measurement
procedure to measure the liquid volume. Here, we use the following
set of values for the components in the gyrator-based circuit $L_{1}=-4.7\mathrm{\:nH}$,
$L_{2}=-4.7\mathrm{\:nH}$, $C_{2}=47\mathrm{\:pF}$, and $R_{g}=50\:\Omega$.
Consider a cylindrical glass with top and bottom metal plates. This
structure can serve as a variable capacitor in which the volume of
filled liquid (or a percentage of a mixture) can change the total
capacitance. A schematic structure for this scenario is illustrated
in Fig. \ref{Fig: Sensor}(a). The designed device includes the gyrator-based
circuit (see Fig. \ref{Fig: Circuit}) where the positive capacitor
on the left side is the cylindrical container with height $d_{2}=3.0142\:\mathrm{cm}$,
of which a height $d_{1}$ is filled with water and the area of metal
plates are $A=100\:\mathrm{cm^{2}}$. Pure water is assumed to have
a relative permittivity of $\varepsilon_{r}=78.7$ at $T=22.0^{\mathrm{o}}\:\mathrm{C}$,
and we neglect losses in this simple case \cite{midi2014broadband}.
Two series variable capacitors model the structure, that the bottom
one has a capacitance $C_{filled}=\varepsilon_{0}\varepsilon_{r}A/d_{1},$and
the top one has a capacitance $C_{empty}=\varepsilon_{0}A/(d_{2}-d_{1})$.
The total capacitance is $C_{total}=C_{filled}C_{empty}/(C_{filled}+C_{empty})$,
which changes when varying the water level. By opening the top inlet,
the height of water increases, so the capacitance value will be increased.
On the contrary, the water's height decreases when opening the bottom
outlet, and the total capacitance value will be decreased. In summary,
a level of water is related to the capacitance, and the perturbation
in the value of capacitance will change a circuit's eigenfrequencies.
Using the explained steps in Section \ref{sec:Lossless-EPD-Condition}
and by solving the eigenvalue problem, the plot of resonance frequency
versus water level percentage for this specific example is illustrated
in Fig. \ref{Fig: Sensor}(b) by the solid blue line. The measuring
scheme is very sensitive near $0\ensuremath{\%}
$ water content. The EPD can be designed for different water contents,
so the frequency variation caused by changes in the water level around
that mentioned level would be very sensitive. We now compare the sensitivity
of the EPD-based scheme with that of a single LC resonator. We consider
an LC resonator with the resonance frequency of $\omega_{0}=\omega_{e}$,
i.e., coincident with one of the EPD systems. We assume that the sensing
capacitor is the same as the one in Fig. \ref{Fig: Sensor}, i.e.,
the same as that considered in the EPD system. The variation in the
resonance frequency by perturbing the capacitance as described above,
i.e., the level of water content, is shown in Fig. \ref{Fig: Sensor}(b)
by the red dashed line. It is clear that the EPD-based bifurcation
in the dispersion diagram, characterized by a square root, dramatically
enhances the circuit's sensitivity compared to the sensitivity of
the single LC resonator to the same capacitance perturbation.

In the proposed scheme for liquid content measurement, we assumed
that the gyrator-based circuit works in the stable region where eigenfrequencies
were purely real. However, when considering the instabilities generated
by losses, one eigenfrequency has a negative imaginary value, as explained
in Section \ref{sec: LossyCircuit}. Consequently, the circuit starts
having growing oscillations. The exponential growth rate can be controlled
in two ways: either by stopping (switching off) the circuit to reach
saturation or by letting it saturate. In this latter case, the gyrator-based
circuit should be designed as a sensor that oscillates. The circuit
can be used to sense physical or chemical parameters changes by measuring
the oscillation frequency variations.

\section{Conclusions}

The second-order EPD with real (degenerate) eigenfrequency in a gyrator-based
circuit is achieved using two unstable series LC resonators coupled
via an ideal gyrator. Both resonators have either negative capacitances
or negative inductances. The resonance frequency of each resonator
(when uncoupled) is purely imaginary, and we have demonstrated that
the EPD frequency can be purely real. However, when losses are considered
in the circuit, the system becomes unstable. We have focused on the
physics of the EPD in this kind of gyrator-based circuit, looking
at enhanced sensitivity when perturbing either the gyration resistance,
a capacitance, or an inductance. In the potential application, the
perturbation in the circuit can be estimated by measuring the shift
of resonance frequencies. The presented results may have significant
implications in sensing technology, security systems, particle monitoring,
and motion sensors.

\section*{Acknowlegment}

This material is based upon work supported by the National Science
Foundation under Grant No. ECCS-1711975 and by AFOSR Grant No. FA9550-19-1-0103.

\bibliographystyle{IEEEtran}


\end{document}